\documentclass[runningheads,10pt]{llncs}

\usepackage[T1]{fontenc}
\usepackage[utf8]{inputenc}
\usepackage{amssymb}
\usepackage{amsfonts}
\usepackage{graphicx}
\usepackage{booktabs}
\usepackage{multirow}
\usepackage{array}
\usepackage{tabularx} 
\usepackage{ragged2e} 
\usepackage[section]{placeins} 
\usepackage{url}
\usepackage{hyperref}
\usepackage{xcolor}
\usepackage{subcaption}
\usepackage[colorinlistoftodos,prependcaption,textsize=tiny]{todonotes}

\usepackage{tikz}
\usetikzlibrary{positioning,arrows.meta}
\usepackage{pgfplots}
\pgfplotsset{compat=1.18}
\pgfplotsset{
  scoreaxis/.style={
    xbar stacked,
    xmin=0, xmax=100,
    width=0.83\textwidth,
    height=0.4\textwidth,
    bar width=6pt,
    enlarge y limits=0.10,
    axis x line*=none,
    axis y line*=none,
    xtick={0,25,50,75,100},
    xticklabels={0\%,25\%,50\%,75\%,100\%},
    xticklabel style={font=\scriptsize},
    ytick=data,
    symbolic y coords={Discovery,Execution,Credential Access,Collection,Initial Access,Lateral Movement,Reconnaissance,Privilege Escalation,Persistence,Defense Evasion,Impact,Total},
    y dir=reverse,
    yticklabel style={font=\scriptsize, text width=2.8cm, align=right},
  },
}

\begin{document}

\title{Decoys Cannot Go Everywhere: Mapping the Deception Surface in MITRE ATT\&CK}
\titlerunning{Mapping the Deception Surface in MITRE ATT\&CK}

\author{Veronica Valeros\inst{1} \and Carlos Catania\inst{2} \and Viliam Lis\'y\inst{1} \and Harm Griffioen\inst{3}}
\authorrunning{Valeros et al.}
\institute{%
Faculty of Electrical Engineering, Czech Technical University in Prague, Czechia\\
\email{valerver@fel.cvut.cz, viliam.lisy@fel.cvut.cz}
\and
School of Engineering, Universidad Nacional de Cuyo, Argentina\\
\email{harpo@ingenieria.uncuyo.edu.ar}
\and
Cyber Threat Intelligence Lab, Delft University of Technology, Netherlands\\
\email{h.j.griffioen@tudelft.nl}%
}

\maketitle

\begin{abstract}
Cyber deception research often assumes that a decoy can be placed wherever there is attacker behavior. This work tests that assumption across MITRE ATT\&CK v18.1. We introduce a four-criterion rubric for infrastructure deception and apply it to all 250 ATT\&CK techniques. The rubric evaluates whether a defender-controlled decoy can be placed, whether an attacker is likely to interact with it, what intelligence that interaction can yield, and whether the interaction reliably indicates malice. The resulting deception surface is sparse: only 80 techniques (32\%) admit a decoy the attacker could plausibly reach. For the remaining 170 techniques, there is no defender-controlled asset in the attacker’s path that can be fabricated as a decoy. Decoy placement across those 80 techniques falls into two patterns we call \textit{Sweep} and \textit{Seek}. In Sweep, the attacker moves broadly through assets in range and encounters the decoy as part of that activity. In Seek, the attacker looks for a specific kind of asset and interacts with a fabricated version of it. These patterns give a simple placement rule: a decoy must either sit on a sweep path or imitate a sought asset. We also show that decoys usually have useful intelligence potential, but whether an attacker interacts with them at all, and whether that interaction reliably indicates malice, both vary. We release the rubric, decision rules, and per-technique assessment as an auditable baseline for future deception research and deployment planning, and show that infrastructure decoys cannot be assumed to apply to all attacker behavior.
\end{abstract}

\section{Introduction}
\label{sec:introduction}
Cyber deception research often assumes that decoys can be placed anywhere in the attack path~\cite{zambiancoProactiveDecoySelection2025,mongardiniSystematicMetaSurveyCyber2026,sayedHoneypotAllocationCyber2023a,alaminCyberDeceptionMetrics2022}. Game theory approaches rely on it to create simplified models of reality to study optimal honeypot placement strategies~\cite{sayedHoneypotAllocationCyber2023a,anwarHoneypotAllocationCyber2022a}. Automated deception frameworks often study engagement actions for different stages of the cyber kill chain without first checking whether the attacker’s technique can realistically be deceived~\cite{sajidSODASystemCyber2021}. Researchers create attack-graph models treating decoys as abstract mathematical entities that can simply be dropped onto any predicted network node or edge~\cite{zambiancoProactiveDecoySelection2025,alaminCyberDeceptionMetrics2022}. Industry efforts such as the MITRE D3FEND framework~\cite{d3fend}, and the MITRE Engage framework~\cite{mitre_engage}, also rely on it. This assumption has not yet been systematically tested. 

We argue that before the community can successfully engineer new deceptive technologies or algorithmically optimize decoy placement, we must first map the \emph{operational deception surface}: where decoys can plausibly be placed in an attacker's path, the interactions they can elicit, and what signal those interactions would yield in practice. A technique \emph{admits a decoy} when carrying it out brings the attacker into contact with an asset the defender can fabricate, instrument, and place in the attacker's path. Mapping this surface is also a precondition for deception-based active defense: a defender cannot decide how to engage an attacker through deception without first knowing where a contact point exists.

This paper proposes to systematically test this assumption through two research questions (RQ): \textit{Which attacker behaviors admit a decoy, and for those that do, what signal would those decoys produce?} (RQ1), and \textit{Why do some attacker behaviors admit a decoy and others do not?} (RQ2). To answer these, we use a concrete catalog of attacker behaviors: MITRE ATT\&CK v18.1~\cite{mitre_attack,al-sadaMITREATTCKState2024}. ATT\&CK is grounded in real-world observations and organizes adversary behavior into 250 techniques across 14 tactics.

To answer RQ1, we developed a four-criterion rubric and applied it to all 250 ATT\&CK v18.1 techniques. The four criteria evaluate: i. whether a decoy can be placed (feasibility), ii. whether an attacker would interact with it (interaction), iii. what signal that interaction would yield (intelligence yield), and iv. whether that signal reliably indicates malice (malice fidelity). To answer RQ2, we analyzed the resulting scores to identify the patterns that separate techniques that admit a decoy from those that do not.

Only 80 (32\%) of the 250 ATT\&CK techniques admit decoys. For the remaining 170, no decoy can be placed to capture the attacker's behavior. Decoy placement across those 80 techniques falls into two patterns we call \textit{Sweep} and \textit{Seek}. In Sweep, the decoy sits in a path where attackers, hitting everything in range, encounter it without seeking it. In Seek, the decoy imitates a sought-after asset, so attackers navigating toward that asset type find it.

\begin{figure}[thb]
    \centering
    \includegraphics[width=\linewidth]{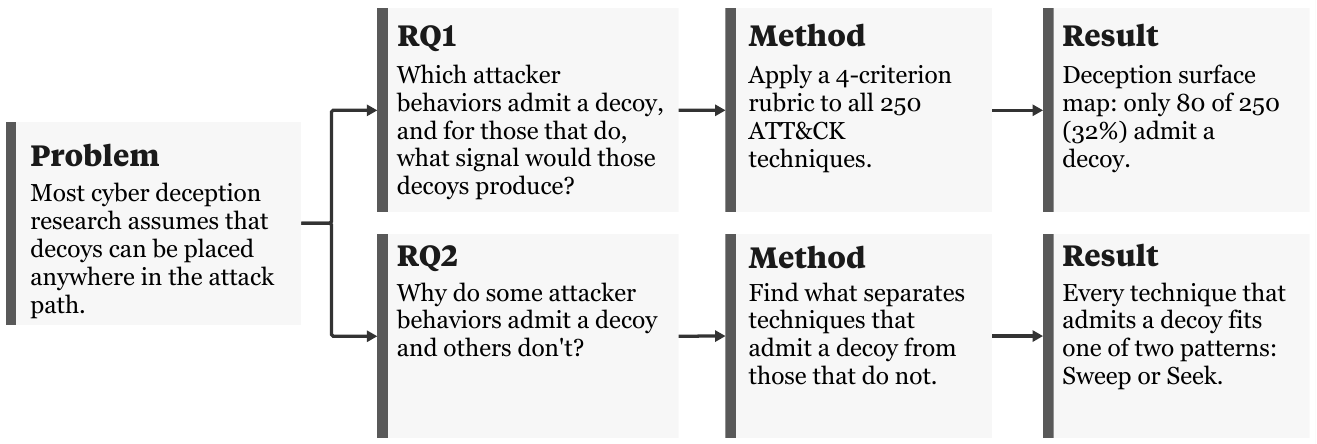}
    \caption{How the paper fits together. Starting with the problem, the figure traces each research question through its method to its result.}
    \label{fig:paper-structure}
\end{figure}

Figure~\ref{fig:paper-structure} lays out the paper at a glance. The initial problem of assuming that a decoy can be placed anywhere in the attack path motivates the two research questions above. Each is answered by a method that yields a concrete result. Together, they show which attacker behaviors a decoy can capture and where it should be placed to work.

The contributions of this paper are:
\begin{itemize}
    \item A four-criterion rubric that enables defenders to evaluate, before deployment, which ATT\&CK techniques admit a decoy and what detection value it provides.
    \item A Sweep--Seek rule that replaces ad-hoc decoy placement. A decoy must lie on a sweep path or imitate an asset the attacker is looking for, or it goes untouched.
    \item An open, ATT\&CK-wide decoy assessment with per-technique scores and rationales, released as a baseline for others to refine and build on.
\end{itemize}

The remainder of this paper is structured as follows: Section~\ref{sec:related} reviews related work. Section~\ref{sec:methodology} defines the rubric, scoring procedure, expert study, and pattern-identification procedure. Section~\ref{sec:results-discussion} presents and interprets the findings. Section~\ref{sec:limitations-future} addresses limitations and future work. Section~\ref{sec:conclusion} concludes.

\section{Related Work}
\label{sec:related}
The concept of placing fake assets in attacker paths to detect and study intrusions dates back to Hollingworth's 1973 work on pseudo-flaws and entrapment modules~\cite{hollingworthEnhancingComputerSystem1973}. The first operational deception followed fifteen years later~\cite{cliffordstollStalkingWilyHacker1988,billcheswickEveningBerferdWhich1992}, and the field has since grown to include honeynets, honeytokens, honeyfiles, decoy credentials, and LLM-driven adaptive systems~\cite{zhangThreeDecadesDeception2021,chouhanCATCHToolAutomatic2025,sladicVelLMesHighinteractionAIbased2025,vasilomanolakisHoneypotdrivenCyberIncident2015}. While the set of decoys has expanded, what attacker behaviors in ATT\&CK admit a decoy remains largely unmapped in the literature~\cite{hanDeceptionTechniquesComputer2018,francoSurveyHoneypotsHoneynets2021,kahlhoferApplicationLayerCyber2024,mongardiniSystematicMetaSurveyCyber2026}.

Substantial work in cyber deception studies decoy placement: given a network, an attack path, or a defender goal, where should decoys be placed to intercept attackers? These approaches use game theory, attack graphs, and related optimization methods to decide where deceptive assets should go~\cite{sayedHoneypotAllocationCyber2023a,anwarHoneypotAllocationCyber2022a,alaminCyberDeceptionMetrics2022}. However, decoys are often treated as objects that can be added to nodes, edges, or paths in the model. While recent works acknowledge that assuming a decoy exists for every technique is a major limitation, they stop short of evaluating whether specific attacker behaviors can actually be deceived in practice~\cite{zambiancoProactiveDecoySelection2025}. Current placement work answers \textit{where} to put a decoy, but not \textit{if} it is possible.

Recent surveys repeatedly note that cyber deception lacks standardized evaluation measures~\cite{mongardiniSystematicMetaSurveyCyber2026,hanDeceptionTechniquesComputer2018,pittmanTaxonomyDynamicHoneypot2020}. Beltrán-López et al.~\cite{beltran-lopezCyberDeceptionTaxonomy2025} show that even the most comprehensive taxonomies frequently omit implementation feasibility. Other areas of cybersecurity have addressed similar problems by making expert judgment explicit through structured scoring or classification schemes. For instance, CVSS~\cite{mell2006common} reduces inconsistent vulnerability scoring through explicit metrics, and STRIDE provides a structured vocabulary for threat modeling~\cite{kohnfelder1999threats}. While recent Systematization of Knowledge (SoK) reviews~\cite{bridgesSoKHoneypotsLLMs2025} demonstrate that the cyber deception community is increasingly capable of applying structured rubrics to systematically evaluate literature, deception defense still lacks an equivalent operational instrument for evaluating implementation feasibility.

Recent work increasingly connects cyber deception to the MITRE ATT\&CK framework to automate active defense and plan how to engage attackers~\cite{chouhanDeceptionTechnologyActive2024,al-sadaMITREATTCKState2024}. Sajid et al.\cite{sajidSODASystemCyber2021} generate dynamic deception playbooks by mapping malware behaviors to specific ATT\&CK tactics. Two industry frameworks, D3FEND~\cite{d3fend} and Engage~\cite{mitre_engage}, both lean on ATT\&CK as the starting point for defensive planning. D3FEND links attacker techniques to relevant defensive responses. Engage organizes deception, denial, and engagement activities that defenders can use to shape attacker behavior. These action-oriented frameworks help defenders decide what to do in response to an adversary technique. However, they do not first ask whether that technique admits a defender-controlled decoy in the attacker’s path. Without that decoy, the deception-based actions these frameworks prescribe have no interaction to respond to.

In summary, prior work has built more types of decoys, studied where to place them, and used ATT\&CK to plan defensive action. What remains missing is an ATT\&CK-wide check of whether each attacker technique admits a defender-controlled decoy in the attacker’s path, and what such a decoy would reveal. This paper addresses that gap.
\section{Methodology}
\label{sec:methodology}
Our methodology has two steps. First, we applied a four-criterion rubric to the ATT\&CK framework to produce a deception surface evaluation that captures where decoys can be placed, what interactions they can elicit, and what signal they can yield (RQ1). Second, we analyzed the assessment to identify patterns that explain why some attacker behaviors admit a decoy and others do not (RQ2). Deciding whether a technique admits a decoy requires expert judgment, so we fixed the decision rules before scoring and applied the same rules to all 250 techniques. This makes each score explicit and keeps the assessment consistent across the matrix.

\subsection{Scope and Definitions}
\label{sec:methodology-scope-defs}
We evaluate one form of deception in ATT\&CK: infrastructure deception, defined as technical decoys under the defender's control deployed within or adjacent to the defender's infrastructure. We exclude information operations, misinformation campaigns, offensive counter-deception, and deception placed outside the defender's infrastructure, because these depend on external actors and conditions the defender cannot control or reliably observe. We score each ATT\&CK technique independently, assuming an adversary capable of the behavior it describes. This means we evaluate what is possible in principle, not whether one specific deployment would succeed.

The unit of analysis is a single ATT\&CK technique. An asset is any resource used to carry out a technique, such as a host, account, credential, or cloud object. A defender-controllable asset is one that the defender can create and manage without the attacker's cooperation. Only these assets can become decoys. An attack path is the chain of techniques an adversary carries out across tactics toward a target. A decoy is reached only if it lies on this path. We therefore score each technique as a potential point on such a path: whether carrying it out brings the attacker into contact with an asset a defender could turn into a decoy. A technique is out of scope when it involves only attacker-internal computation or assets the defender cannot observe or control, such as obfuscating their own code or modifying timestamps on a host they already control. A full list of asset types is given in Appendix~\ref{app:assets}.

\subsection{Evaluation Criteria}
\label{sec:methodology-rubric-scoring}

\begin{figure}[!htbp]
    \centering
    \includegraphics[width=\columnwidth]{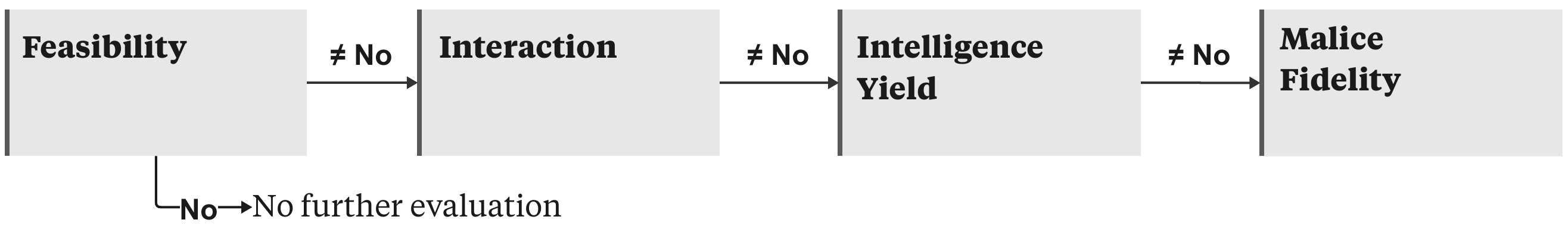}
    \caption{The four-criterion rubric. Each criterion is assessed only if the previous did not score No; the first No ends the evaluation.}
    \label{fig:evaluation-criteria}
\end{figure}

To answer RQ1, we applied a four-criterion rubric to all 250 ATT\&CK v18.1 techniques. The rubric comprises four criteria: feasibility, interaction, intelligence yield, and malice fidelity. Feasibility acts as a gate: if a technique offers no defender-controllable asset to mimic, the remaining criteria are not assessed. The criteria are then applied in sequence (Figure~\ref{fig:evaluation-criteria}), each assuming the previous one held. Scores are reported per criterion, each reflecting the best case a well-instrumented decoy could plausibly produce. Each criterion is described below:
\begin{itemize}
    \item \textbf{Feasibility:} Does this technique rely on any asset that could realistically be mimicked as a decoy?
    \item \textbf{Interaction:} Does this technique make it likely that an attacker will interact with the decoy? Interaction means the attacker actively engages with the decoy, producing observable data.
    \item \textbf{Intelligence Yield:} Does interaction with the decoy have the potential to yield strategic, operational, tactical, or technical intelligence? The actual yield depends on deployment design and measurement, which are out of scope here.
    \item \textbf{Malice Fidelity:} If an attacker interacts with the decoy, how reliably does that signal malicious intent rather than benign activity? 
\end{itemize}

\subsection{Scoring Procedure}
\label{sec:methodology-scale}

All four criteria use a four-point forced-choice scale with no neutral midpoint. The four levels are \textbf{Yes}, \textbf{Mostly Yes}, \textbf{Mostly No}, and \textbf{No}. Removing the neutral option forces a clear judgment on every criterion and avoids a default \textit{maybe} or \textit{it depends}, which is a real risk given how much ATT\&CK relies on expert interpretation.

Each criterion has a guide describing what each of the four levels means, fixed before scoring. Table~\ref{tab:rubric-combined} presents the complete guides. The levels form an ordered scale, from Yes down to No. Where evidence fits two adjacent levels equally, we assign the lower one (closer to No).

\begin{table}[tb]
\centering
\caption{Scoring guide for the four rubric criteria.}
\label{tab:rubric-combined}
    \begingroup\fontsize{7pt}{9pt}\selectfont 
    \begin{tabularx}{\columnwidth}{>{\RaggedRight\arraybackslash}p{1.35cm}>{\RaggedRight\arraybackslash}X>{\RaggedRight\arraybackslash}X>{\RaggedRight\arraybackslash}X>{\RaggedRight\arraybackslash}X}
        \toprule
        \textbf{Score} & \textbf{Feasibility} & \textbf{Interaction} & \textbf{Intelligence Yield} & \textbf{Malice Fidelity} \\
        \midrule
        \textbf{Yes} &
            The defender can fully fabricate and control the target asset as a decoy, and it responds convincingly to attacker actions. &
            
            The technique naturally leads attackers to the decoy. Interaction follows as a direct consequence of the technique. &
            
            Interaction directly yields strategic, operational, tactical, or technical intelligence attributable to the decoy. &
            
            Legitimate interaction is not expected by design. The only plausible trigger is an attacker action, so the false-positive rate is near zero. \\
        \addlinespace[6pt]
        \textbf{Mostly Yes} &
            The target asset can be mimicked but it is hard to make convincing. May not withstand close scrutiny. &
            
            Interaction is likely but not certain, depending on the decoy's positioning, configuration, or the attacker’s tools. &
            
            Interaction yields intelligence, but only after correlation with other data, added context, or further analysis. &
            
            Interaction strongly indicates malice. A small set of benign activities could trigger it, but these cases are identifiable and filterable. \\
        \addlinespace[6pt]
        \textbf{Mostly No} &
            The target asset can only be partially mimicked as a decoy. It is difficult to simulate convincingly and only works in limited conditions. &
            
            Interaction is possible but unlikely, requiring attacker-specific knowledge, unusual timing, or atypical choices. &
            
            Interaction produces some data, but it is too generic or ambiguous to be meaningful without significant further analysis. &
            
            Interaction may indicate malice, but many triggers are benign or ambiguous. Telling them apart is complex, so the signal is useful but not standalone. \\
        \addlinespace[6pt]
        \textbf{No} &
            The technique has no defender-controllable target asset that can be fabricated and operated as a decoy. &
            
            No plausible attacker path to the decoy exists. An attacker following the technique would not be expected to interact with this decoy. &
            
            No intelligence yield. The observable data gives no insight into the attacker's behavior, identity, or intent. &
            
            Benign activity routinely triggers this decoy. Interaction does not distinguish an attacker. \\
        \bottomrule
    \end{tabularx}
    \endgroup
\end{table}

All 250 techniques were scored by a single evaluator, in the order they appear within each tactic, using the following steps:

\begin{enumerate}
    \item Read the ATT\&CK technique description and procedure examples, which show how the technique has been observed in real-world attacks.
    \item Identify the assets involved in the technique. For each asset, determine whether the defender can fabricate, control, or instrument it.
    \item Score each criterion sequentially following the scoring guide.
    \item Record criteria rating and rationale for later post-processing and analysis.
\end{enumerate}

As an example, consider Brute Force (T1110). Feasibility: a login service can be turned into a decoy, so yes. Interaction: a login decoy naturally attracts credential-guessing tools that hit every reachable account, giving yes. Intelligence yield: a hit reveals solid tactical and technical intelligence directly attributable to the decoy, so yes. Malice fidelity: automated scanners and misconfigured tools can generate some authentication noise against open services, but the signal stays high quality after basic filtering, giving Mostly Yes.

These scores capture whether a technique has the preconditions for deception, not whether a given deployment will succeed. A high score means the opportunity exists in principle. Whether a real deployment works depends on implementation and operational context.

\subsection{Expert Study}
\label{sec:methodology-expert-study}

After our evaluation we ran a small expert study with two aims. First, we wanted to see whether people other than the lead author could apply the rubric from written instructions. Second, we wanted to measure how closely experts agreed when scoring the same techniques. We recruited eight experts through convenience sampling from our professional networks. The experts work in cyber deception, threat intelligence, and security operations.

Each expert independently scored five ATT\&CK techniques using the four rubric criteria: T1110 (Brute Force), T1132 (Data Encoding), T1213 (Data from Information Repositories), T1218 (System Binary Proxy Execution), and T1485 (Data Destruction). We selected these techniques to include both clear and ambiguous cases. Sessions were kept under 45 minutes. Since each technique took 5 to 10 minutes to score, we limited the study to five techniques per expert. This produced 20 scores per expert, for a total of 160 expert scores.

Experts gave verbal consent for audio recording and anonymized, aggregated reporting. During a one-on-one call, each expert received a one-page rubric summary and the official ATT\&CK descriptions. Experts selected scores independently while explaining their reasoning aloud. After scoring, we held a short debrief to collect feedback on ambiguous wording, difficult decisions, and parts of the rubric that were hard to apply.

We analyzed the study in two ways. First, we reviewed the transcripts to see whether experts could apply the rubric and to identify where they interpreted a criterion or technique differently. Second, we compared the scores numerically. Since the goal was to test whether the rubric produces comparable judgments across raters, we also included the lead author's existing scores for the same five techniques. We treated the author as one more rater, not as the answer key. This gave us nine raters in total and a complete set of ratings: all nine raters scored all five techniques on all four criteria.

Because the rubric scores are ordered, we measured both exact agreement and closeness. Exact agreement asks whether raters gave the same score. Closeness asks how far apart their scores were, since a disagreement between Yes and Mostly Yes is smaller than a disagreement between Yes and No. We report exact agreement, ordered agreement, and Krippendorff's ordinal alpha ($\alpha$), which is a standard agreement measure for multiple raters on an ordered scale~\cite{hayesAnsweringCallStandard2007}.

\subsection{Identifying Placement Patterns}
\label{sec:methodology-patterns}

To answer RQ2, we split techniques into two sets according to their Feasibility score: those that scored No, and those that did not. For each set, we reviewed the defender-controllable assets we identified during the feasibility scoring and analyzed how an attacker would interact with a decoy, or why no such interaction was possible. We created groups based on the identified type of interaction. The resulting groups are reported in Section~\ref{sec:results}.

\section{Results and Discussion}
\label{sec:results-discussion}
\label{sec:results} 
\label{sec:discussion} 

This section presents and interprets the findings. We first map the deception surface across ATT\&CK and show that decoy feasibility is sparse. We then explain why some techniques admit a decoy, and others do not. We close with the results of the expert study and a discussion about the scoring agreement.

\subsection{The deception surface across ATT\&CK}
\label{sec:results-surface}
The deception surface across ATT\&CK is sparse. Of the 250 ATT\&CK v18.1 techniques, only 80 (32\%) scored anything other than No on Feasibility. For the remaining 170 techniques (68\%), no defender-controlled decoy can be placed in the attacker’s path. Among the 80 techniques that admit a decoy, 58 scored Yes on Feasibility, 12 scored Mostly Yes, and 10 scored Mostly No. This means that, across ATT\&CK, deception can be positioned against at most 32\% of cataloged techniques.

\begin{figure}[t]
    \centering
    \includegraphics[width=\linewidth]{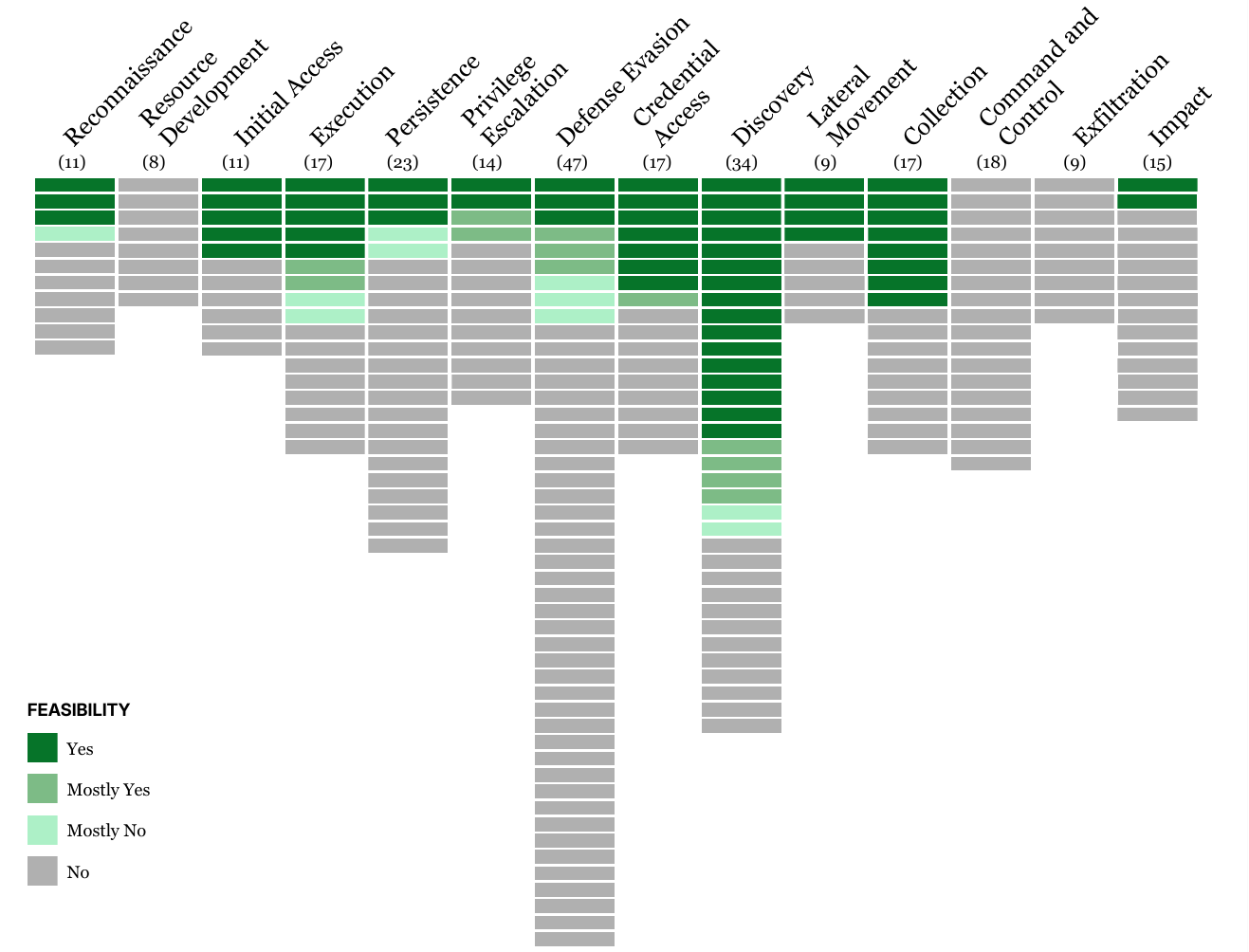}
    \caption{Deception surface across the ATT\&CK matrix. Techniques that admit a decoy are those with Feasibility $\neq$ No.}
    \label{fig:viability-map}
\end{figure}

The surface is also uneven. Techniques that admit a decoy are concentrated in a few tactics rather than spread evenly across the matrix. The distribution is shown in Table~\ref{tab:summary}, and the deception surface is shown in Figure~\ref{fig:viability-map}. Five tactics contain 52 of the 80 techniques that admit a decoy: Discovery (22 of 34), Execution (9 of 17), Credential Access (8 of 17), Collection (8 of 17), and Initial Access (5 of 11). Discovery has the largest count, with 22 techniques. At the other end, three tactics have no techniques that admit a decoy: Resource Development, Command and Control, and Exfiltration.

\begin{table*}[!htbp]
\centering
\caption{Techniques that admit a decoy (Feasibility $\neq$ No), by ATT\&CK tactic, ordered from highest to lowest share of techniques.}
\label{tab:summary}
\begingroup\fontsize{7pt}{9pt}\selectfont 
\setlength{\tabcolsep}{5pt}
\renewcommand{\arraystretch}{1.1}
\begin{tabular}{llrrrrrr}
\toprule
\multicolumn{3}{c}{\textbf{ATT\&CK Tactic}} &
\multicolumn{3}{c}{\textbf{Feasibility score}} &
\multicolumn{2}{c}{\textbf{Admits a decoy}} \\
\cmidrule(lr){1-3}\cmidrule(lr){4-6}\cmidrule(lr){7-8}
ID & Name & Total & Yes & Mostly Yes & Mostly No & Count & \% \\
\midrule
TA0007 & Discovery             & 34 & 16 & 4 & 2 & 22 & 65\% \\
TA0002 & Execution             & 17 &  5 & 2 & 2 &  9 & 53\% \\
TA0006 & Credential Access     & 17 &  7 & 1 & 0 &  8 & 47\% \\
TA0009 & Collection            & 17 &  8 & 0 & 0 &  8 & 47\% \\
TA0001 & Initial Access        & 11 &  5 & 0 & 0 &  5 & 45\% \\
TA0008 & Lateral Movement      &  9 &  4 & 0 & 0 &  4 & 44\% \\
TA0043 & Reconnaissance        & 11 &  3 & 0 & 1 &  4 & 36\% \\
TA0004 & Privilege Escalation  & 14 &  2 & 2 & 0 &  4 & 29\% \\
TA0003 & Persistence           & 23 &  3 & 0 & 2 &  5 & 22\% \\
TA0005 & Defense Evasion       & 47 &  3 & 3 & 3 &  9 & 19\% \\
TA0040 & Impact                & 15 &  2 & 0 & 0 &  2 & 13\% \\
TA0042 & Resource Development  &  8 &  0 & 0 & 0 &  0 &  0\% \\
TA0011 & Command and Control   & 18 &  0 & 0 & 0 &  0 &  0\% \\
TA0010 & Exfiltration          &  9 &  0 & 0 & 0 &  0 &  0\% \\
\midrule
\textbf{Total} & & \textbf{250} & \textbf{58} & \textbf{12} & \textbf{10} & \textbf{80} & \textbf{32\%} \\
\bottomrule
\end{tabular}
\endgroup
\end{table*}

\subsection{Why some techniques admit decoys}
\label{sec:results-why}

The split between the 80 techniques that admit a decoy and the 170 that do not comes down to one condition: whether using the technique makes the attacker interact with an asset the defender controls. Every technique that admits a decoy drives the attacker toward something the defender can fabricate, such as a login service, credential, share, host, mailbox, or file. Credential guessing seeks accounts, ransomware looks for files, enumeration tries to find reachable services. In each case, the defender has a place to deploy a fake.

The 170 techniques that do not admit a decoy fail this condition in one of two ways. First, some techniques happen outside defender-controlled space, or depend on attacker-controlled infrastructure. The defender has no asset to place where the attacker will reach. This explains the three tactics with no techniques that admit a decoy: Resource Development (8 techniques), Command and Control (18), and Exfiltration (9). Second, some techniques involve the attacker acting only on objects they already control on the compromised host, such as obfuscating code, modifying timestamps, disabling security tools, scheduling tasks, or exploiting local vulnerabilities. There is nowhere for a decoy to go because the attacker never reaches for a defender-controllable resource. This covers most techniques that do not admit a decoy in Defense Evasion (38 of 47), Persistence (18 of 23), and Privilege Escalation (10 of 14), and also appears in smaller numbers across the remaining tactics.

This explains the shape of the surface illustrated in Figure~\ref{fig:viability-map}. The empty tactics are stages where, under our scope of infrastructure deception, the attacker behavior does not create a place for a defender-controlled decoy. This does not mean deception has no role there. It suggests that those stages may be better suited to persona-based or narrative deception, where the goal is to shape what the attacker believes or chooses rather than to interact with a decoy asset.

\subsection{Placement patterns: Sweep and Seek}
\label{sec:sweep-seek}
Among the 80 techniques that admit a decoy, our grouping produced two patterns. In the first pattern, which we call \textbf{Sweep}, the attacker hits everything in range and encounters the decoy as a side effect. File collection, network scanning, credential spraying, ransomware, and wipers all follow this pattern. The decoy does not need to be specifically sought. It needs to sit where the bulk activity passes through. In the second pattern, which we call \textbf{Seek}, the attacker is looking for a specific kind of asset, either to read it or to act on it, and interacts with a fabricated version of that asset. Credentials, services, shares, hosts, and mailboxes are all assets an attacker may seek and a defender can fake. Every technique that admits a decoy falls into one of these two patterns.

The two patterns also give a placement rule. A decoy works only if it sits on a sweep path or imitates a sought asset. A Sweep decoy must be placed where broad activity will pass through it. A Seek decoy must look like the asset the attacker is trying to find. A decoy placed outside both paths is unlikely to be reached, regardless of how convincing it is.

For defenders, this makes placement depend on the technique being targeted. If the technique sweeps, the decoy must sit among the assets that will be swept. If the technique seeks, the decoy must resemble the asset being sought. This also explains why decoys placed without regard to attacker behavior may go untouched: they sit outside both paths.

\subsection{What signal do decoys produce?}
\label{sec:results-signal}
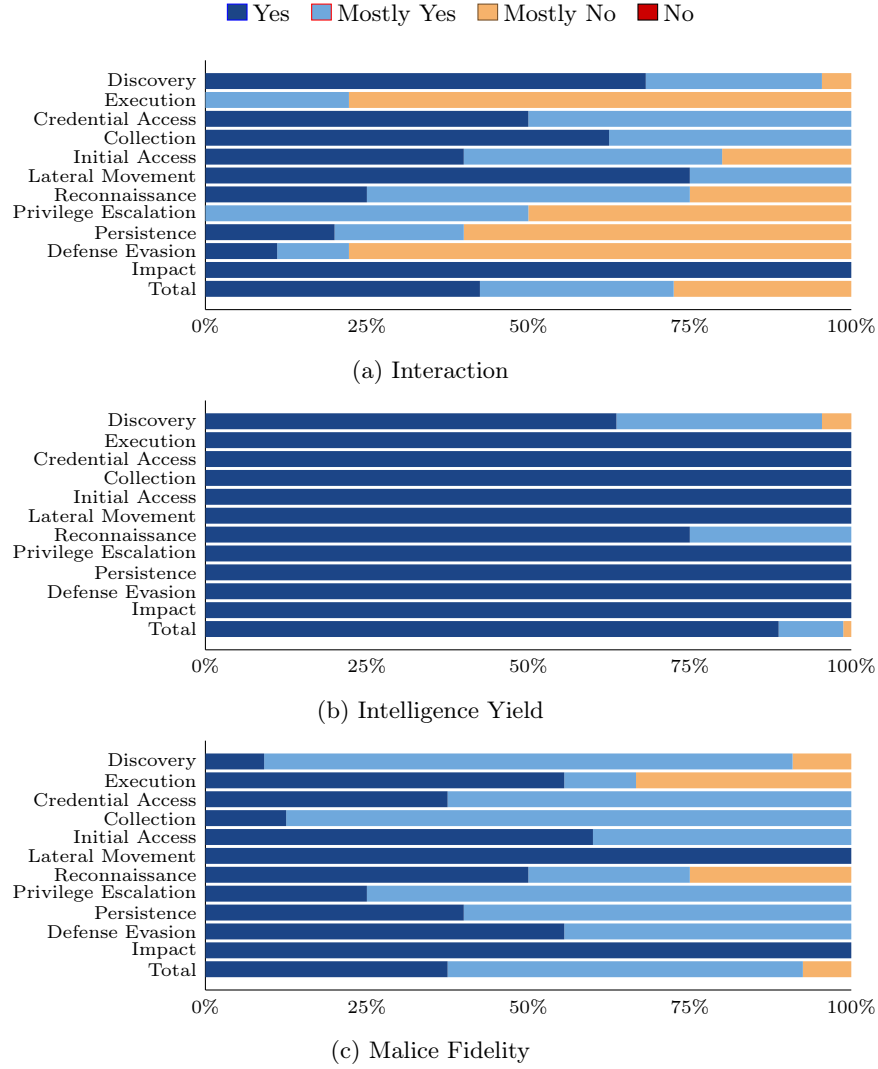
\begin{figure*}[!htbp]
    \centering
    \begingroup
    \small
    \definecolor{ScoreYes}{HTML}{1C4587}
    \definecolor{ScoreMostlyYes}{HTML}{6FA8DC}
    \definecolor{ScoreMostlyNo}{HTML}{F6B26B}
    \definecolor{ScoreNo}{HTML}{CC0000}

    \begin{subfigure}[t]{\textwidth}
        \centering
        \begin{tikzpicture}
        \begin{axis}[
            scoreaxis,
            legend style={
                at={(0.02,1.10)},
                anchor=south west,
                legend columns=4,
                draw=none,
                /tikz/every even column/.append style={column sep=6pt},
                font=\small,
            },
        ]
            \addplot+[fill=ScoreYes,draw=none] coordinates {
                (68.18,Discovery) (0,Execution) (50,Credential Access) (62.5,Collection) (40,Initial Access) (75,Lateral Movement) (25,Reconnaissance) (0,Privilege Escalation) (20,Persistence) (11.11,Defense Evasion) (100,Impact) (42.5,Total)
            };
            \addlegendentry{Yes}

            \addplot+[fill=ScoreMostlyYes,draw=none] coordinates {
                (27.27,Discovery) (22.22,Execution) (50,Credential Access) (37.5,Collection) (40,Initial Access) (25,Lateral Movement) (50,Reconnaissance) (50,Privilege Escalation) (20,Persistence) (11.11,Defense Evasion) (0,Impact) (30,Total)
            };
            \addlegendentry{Mostly Yes}

            \addplot+[fill=ScoreMostlyNo,draw=none] coordinates {
                (4.55,Discovery) (77.78,Execution) (0,Credential Access) (0,Collection) (20,Initial Access) (0,Lateral Movement) (25,Reconnaissance) (50,Privilege Escalation) (60,Persistence) (77.78,Defense Evasion) (0,Impact) (27.5,Total)
            };
            \addlegendentry{Mostly No}

            \addplot+[fill=ScoreNo,draw=none] coordinates {
                (0,Discovery) (0,Execution) (0,Credential Access) (0,Collection) (0,Initial Access) (0,Lateral Movement) (0,Reconnaissance) (0,Privilege Escalation) (0,Persistence) (0,Defense Evasion) (0,Impact) (0,Total)
            };
            \addlegendentry{No}

            \node[anchor=west,font=\scriptsize] at (axis cs:101,Discovery) {n=22};
            \node[anchor=west,font=\scriptsize] at (axis cs:101,Execution) {n=9};
            \node[anchor=west,font=\scriptsize] at (axis cs:101,Credential Access) {n=8};
            \node[anchor=west,font=\scriptsize] at (axis cs:101,Collection) {n=8};
            \node[anchor=west,font=\scriptsize] at (axis cs:101,Initial Access) {n=5};
            \node[anchor=west,font=\scriptsize] at (axis cs:101,Lateral Movement) {n=4};
            \node[anchor=west,font=\scriptsize] at (axis cs:101,Reconnaissance) {n=4};
            \node[anchor=west,font=\scriptsize] at (axis cs:101,Privilege Escalation) {n=4};
            \node[anchor=west,font=\scriptsize] at (axis cs:101,Persistence) {n=5};
            \node[anchor=west,font=\scriptsize] at (axis cs:101,Defense Evasion) {n=9};
            \node[anchor=west,font=\scriptsize] at (axis cs:101,Impact) {n=2};
            \node[anchor=west,font=\scriptsize] at (axis cs:101,Total) {n=80};
        \end{axis}
        \end{tikzpicture}
        \caption{Interaction}
    \end{subfigure}

    \vspace{4pt}

    \begin{subfigure}[t]{\textwidth}
        \centering
        \begin{tikzpicture}
        \begin{axis}[scoreaxis]
            \addplot+[fill=ScoreYes,draw=none] coordinates {
                (63.64,Discovery) (100,Execution) (100,Credential Access) (100,Collection) (100,Initial Access) (100,Lateral Movement) (75,Reconnaissance) (100,Privilege Escalation) (100,Persistence) (100,Defense Evasion) (100,Impact) (88.75,Total)
            };
            \addplot+[fill=ScoreMostlyYes,draw=none] coordinates {
                (31.82,Discovery) (0,Execution) (0,Credential Access) (0,Collection) (0,Initial Access) (0,Lateral Movement) (25,Reconnaissance) (0,Privilege Escalation) (0,Persistence) (0,Defense Evasion) (0,Impact) (10,Total)
            };
            \addplot+[fill=ScoreMostlyNo,draw=none] coordinates {
                (4.55,Discovery) (0,Execution) (0,Credential Access) (0,Collection) (0,Initial Access) (0,Lateral Movement) (0,Reconnaissance) (0,Privilege Escalation) (0,Persistence) (0,Defense Evasion) (0,Impact) (1.25,Total)
            };
            \addplot+[fill=ScoreNo,draw=none] coordinates {
                (0,Discovery) (0,Execution) (0,Credential Access) (0,Collection) (0,Initial Access) (0,Lateral Movement) (0,Reconnaissance) (0,Privilege Escalation) (0,Persistence) (0,Defense Evasion) (0,Impact) (0,Total)
            };
        \end{axis}
        \end{tikzpicture}
        \caption{Intelligence Yield}
    \end{subfigure}

    \vspace{4pt}

    \begin{subfigure}[t]{\textwidth}
        \centering
        \begin{tikzpicture}
        \begin{axis}[scoreaxis]
            \addplot+[fill=ScoreYes,draw=none] coordinates {
                (9.09,Discovery) (55.56,Execution) (37.5,Credential Access) (12.5,Collection) (60,Initial Access) (100,Lateral Movement) (50,Reconnaissance) (25,Privilege Escalation) (40,Persistence) (55.56,Defense Evasion) (100,Impact) (37.5,Total)
            };
            \addplot+[fill=ScoreMostlyYes,draw=none] coordinates {
                (81.82,Discovery) (11.11,Execution) (62.5,Credential Access) (87.5,Collection) (40,Initial Access) (0,Lateral Movement) (25,Reconnaissance) (75,Privilege Escalation) (60,Persistence) (44.44,Defense Evasion) (0,Impact) (55,Total)
            };
            \addplot+[fill=ScoreMostlyNo,draw=none] coordinates {
                (9.09,Discovery) (33.33,Execution) (0,Credential Access) (0,Collection) (0,Initial Access) (0,Lateral Movement) (25,Reconnaissance) (0,Privilege Escalation) (0,Persistence) (0,Defense Evasion) (0,Impact) (7.5,Total)
            };
            \addplot+[fill=ScoreNo,draw=none] coordinates {
                (0,Discovery) (0,Execution) (0,Credential Access) (0,Collection) (0,Initial Access) (0,Lateral Movement) (0,Reconnaissance) (0,Privilege Escalation) (0,Persistence) (0,Defense Evasion) (0,Impact) (0,Total)
            };
        \end{axis}
        \end{tikzpicture}
        \caption{Malice Fidelity}
    \end{subfigure}

    \endgroup
    \caption{Downstream score breakdown by tactic (percent of techniques), for the 80 techniques that admit a decoy (Feasibility $\neq$ No). The y-axis tactics follow the same order as Table~\ref{tab:downstream-tactic}.}
    \label{fig:downstream-by-tactic}
\end{figure*}

For the 80 techniques that admit a decoy, we examined what happens after placement: whether attackers are likely to interact with the decoy, whether that interaction yields intelligence, and whether it reliably indicates malice. The breakdown of these scores per tactic is shown in Table~\ref{tab:downstream-tactic}. A visual representation of these scores is also shown in Figure~\ref{fig:downstream-by-tactic}.

\begin{table*}[htbp]
\centering
\caption{Downstream score breakdown by ATT\&CK tactic for the 80 techniques that admit a decoy (Feasibility $\neq$ No). Downstream scores are Interaction, Intelligence Yield, and Malice Fidelity. MY = Mostly Yes, MN = Mostly No.}
\label{tab:downstream-tactic}
\begingroup\fontsize{7pt}{9pt}\selectfont 
\setlength{\tabcolsep}{3.8pt}
\renewcommand{\arraystretch}{0.95}
\begin{tabular}{llrrrrrrrrrrrr}
\toprule
\multicolumn{2}{c}{\textbf{ATT\&CK Tactic}} &
\multicolumn{4}{c}{\textbf{Interaction}} &
\multicolumn{4}{c}{\textbf{Intelligence Yield}} &
\multicolumn{4}{c}{\textbf{Malice Fidelity}} \\
\cmidrule(lr){1-2}\cmidrule(lr){3-6}\cmidrule(lr){7-10}\cmidrule(lr){11-14}
ID & Name & Yes & MY & MN & No & Yes & MY & MN & No & Yes & MY & MN & No \\
\midrule
TA0007 & Discovery             & 15 & 6 & 1 & 0 & 14 & 7 & 1 & 0 & 2 & 18 & 2 & 0 \\
TA0002 & Execution             &  0 & 2 & 7 & 0 &  9 & 0 & 0 & 0 & 5 &  1 & 3 & 0 \\
TA0006 & Credential Access     &  4 & 4 & 0 & 0 &  8 & 0 & 0 & 0 & 3 &  5 & 0 & 0 \\
TA0009 & Collection            &  5 & 3 & 0 & 0 &  8 & 0 & 0 & 0 & 1 &  7 & 0 & 0 \\
TA0001 & Initial Access        &  2 & 2 & 1 & 0 &  5 & 0 & 0 & 0 & 3 &  2 & 0 & 0 \\
TA0008 & Lateral Movement      &  3 & 1 & 0 & 0 &  4 & 0 & 0 & 0 & 4 &  0 & 0 & 0 \\
TA0043 & Reconnaissance        &  1 & 2 & 1 & 0 &  3 & 1 & 0 & 0 & 2 &  1 & 1 & 0 \\
TA0004 & Privilege Escalation  &  0 & 2 & 2 & 0 &  4 & 0 & 0 & 0 & 1 &  3 & 0 & 0 \\
TA0003 & Persistence           &  1 & 1 & 3 & 0 &  5 & 0 & 0 & 0 & 2 &  3 & 0 & 0 \\
TA0005 & Defense Evasion       &  1 & 1 & 7 & 0 &  9 & 0 & 0 & 0 & 5 &  4 & 0 & 0 \\
TA0040 & Impact                &  2 & 0 & 0 & 0 &  2 & 0 & 0 & 0 & 2 &  0 & 0 & 0 \\
\midrule
\textbf{Total} & & \textbf{34} & \textbf{24} & \textbf{22} & \textbf{0} & \textbf{71} & \textbf{8} & \textbf{1} & \textbf{0} & \textbf{30} & \textbf{44} & \textbf{6} & \textbf{0} \\
\bottomrule
\end{tabular}
\endgroup
\end{table*}

On \textbf{Interaction}, 34 techniques scored Yes, 24 scored Mostly Yes, and 22 scored Mostly No. This means that even when a decoy can be placed, attacker interaction is not always equally likely. The reason is placement: some techniques naturally drive attackers toward the decoy, while others depend on the decoy being in the right place, having the right configuration, or matching the attacker’s tools. For defenders, the takeaway is that Feasibility alone is not enough. A decoy may be possible, but still unlikely to be touched unless it is placed where the technique actually leads the attacker.

On \textbf{Intelligence Yield}, 71 techniques scored Yes, 8 scored Mostly Yes, and only 1 scored Mostly No. This shows that yield is rarely the limiting factor once a decoy admits interaction. When an attacker interacts with a decoy, that interaction usually reveals something useful: the asset they wanted, the tool they used, the action they attempted, or the stage of the attack they were in. While intelligence yield depends on how is implemented, for defenders, the harder problem is getting the attacker to interact with the decoy in the first place.

On \textbf{Malice Fidelity}, 30 techniques scored Yes, 44 scored Mostly Yes, and 6 scored Mostly No. This is where the strongest trade-off appears. Discovery (TA0007) admits the most decoys of any tactic, with 22 techniques, but only 2 of those 22 score Yes on Malice Fidelity because legitimate users also enumerate files, accounts, and services. Lateral Movement (TA0008) and Impact (TA0040) sit at the other end: every technique that admits a decoy in those tactics scores Yes on Malice Fidelity, but together they account for only six techniques. For defenders, the right placement strategy depends on the goal. Broad coverage comes from decoys in busy places, where hits need correlation and filtering before they can be treated as malicious. High Malice Fidelity comes from decoys in places legitimate users should not touch, but those opportunities are fewer.

\subsection{Expert study: rubric use and agreement}
\label{sec:results-expert-study}

\subsubsection{Can experts apply the rubric?}
\label{sec:results-expert-apply}

All eight experts applied the rubric end-to-end. Every expert produced a score and a spoken rationale for all five techniques, giving the full 160 expected expert scores. This answers the first aim of the study: experts other than the original author could use the rubric from written instructions.

The transcript review showed that experts understood the overall task, but boundary cases were difficult to evaluate. The most common difficulty was separating a decoy that could exist from a decoy the attacker would actually touch. Experts often agreed that a defender could fabricate some asset, but disagreed on whether the attacker path would realistically lead to it. This was especially visible for Interaction.

The transcripts also showed three other sources of disagreement. First, experts differed on how convincing a partial decoy needed to be to count as feasible. Second, some experts treated a decoy hit as useful intelligence if it revealed tools, wordlists, infrastructure, or intent, while others separated this from mere detection and scored Intelligence Yield more conservatively. Third, some experts treated destructive actions or access to sensitive repositories as clear evidence of malice, while others considered benign administrative activity or internal noise as possible explanations. Some ATT\&CK techniques cover different situations under the same technique name. For example, Brute Force can mean trying passwords against a live service or cracking hashes offline. These cases made it harder to choose a single score.
\subsubsection{How closely did experts agree?}
\label{sec:results-expert-agreement}

Table~\ref{tab:agreement} reports agreement across nine raters (see Section~\ref{sec:methodology-expert-study}). Raters picked the exact same level between 28\% and 44\% of the time. When we also counted near-misses where two raters were one step apart, such as Yes and Mostly Yes, agreement
rose to between 67\% and 76\%. In other words, raters rarely landed on the identical answer, but they were usually close, and almost never gave opposite answers. The chance-corrected score (Krippendorff's $\alpha$) was low for every criterion, which we explain next.

\begin{table}[t]
\centering
\caption{Agreement across nine raters on five ATT\&CK techniques. Exact agreement shows how often raters chose the same score. Ordered agreement also gives credit when scores were close, such as Yes versus Mostly Yes. $\alpha$ is Krippendorff's ordinal alpha.}
\label{tab:agreement}
\begin{tabular}{lccc}
\toprule
Criterion & Exact & Ordered & $\alpha$ \\
\midrule
Feasibility         & 0.42 & 0.72 & \phantom{$-$}0.36 \\
Interaction         & 0.28 & 0.67 & $-$0.06 \\
Intelligence Yield  & 0.39 & 0.73 & \phantom{$-$}0.08 \\
Malice Fidelity     & 0.44 & 0.76 & \phantom{$-$}0.12 \\
\bottomrule
\end{tabular}
\end{table}

The low $\alpha$ values do not mean that raters gave opposite judgments. They reflect what this statistic is designed to measure. Unlike the first two columns, Krippendorff's $\alpha$ corrects for agreement that could happen by chance. In our study, this correction is hard to interpret because the sample is small and many scores fall in the upper levels of the scale. Under those conditions, chance-corrected scores can be low, or even slightly negative, even when raters mostly chose neighboring levels. We therefore interpret the expert study as showing close but not identical use of the rubric, rather than strong chance-corrected agreement.

Taken together, the study supports the rubric as usable, but not settled. Experts could apply it from written instructions and usually reached nearby judgments. The study also identifies concrete refinements: sharpen the difference between reachability and interaction, clarify what counts as intelligence rather than detection alone, make Malice Fidelity more explicit about benign administrative activity, and note when ATT\&CK technique definitions combine scenarios that may need separate scoring.

\section{Limitations and Future Work}
\label{sec:limitations-future}

The main limitation of this work is that the full ATT\&CK assessment was conducted by a single primary scorer. While we attempted to reduce this risk by using fixed guided scoring, reviewing ambiguous cases, and running a small expert study, this remains a limitation. The deception surface map should therefore be treated as a first systematic assessment that invites discussion, not as a final truth. MITRE ATT\&CK continues to evolve and new technology allows for new assets and new possible decoys. Future work should update the deception surface map as ATT\&CK evolves, involve more raters, and use community review to refine the scoring rules. LLM-based judges or critics may also be useful as an additional consistency check, but they should support rather than replace expert review.

A second limitation is that the rubric evaluates what is possible in principle, not what succeeds in a particular deployment. A technique may admit a decoy, but the outcome still depends on how the decoy is implemented, where it is placed, and whether the attacker notices it. Future work should apply the rubric to deployed deception systems and compare the scores with observed deployment outcomes.

The analysis also showed that many ATT\&CK techniques share the same underlying decoy asset. For example, a honey file that gets accessed can be used to capture more than one attacker behavior. Future work should develop an asset-centered assessment alongside the technique-centered one. The technique map shows which attacker behaviors a decoy can intercept. The asset view can show what defenders need to deploy.

\section{Conclusion}
\label{sec:conclusion}

Cyber deception often assumes that a decoy can be placed wherever there is attacker behavior. We tested that assumption across ATT\&CK v18.1. Using a fixed four-criterion rubric applied to all 250 techniques, we found that only 80 techniques (32\%) admit a defender-controlled decoy that the attacker could plausibly reach. For the remaining 170 techniques, there is no defender-controlled asset in the attacker’s path that can be fabricated as a decoy.

This boundary is not only a scoring result. It reflects where the attack takes place. A decoy works only when the attacker reaches for something the defender can fabricate, instrument, and observe. That condition appears in some tactics and is absent from others under our scope of infrastructure deception. This explains why decoy opportunities are concentrated in tactics such as Discovery, Execution, Credential Access, and Collection, and absent from Resource Development, Command and Control, and Exfiltration.

For the techniques that admit a decoy, the attacker reaches it in one of two ways. In Sweep, the attacker moves broadly through everything in range and encounters the decoy as part of that activity. In Seek, the attacker looks for a specific kind of asset and interacts with a fabricated version of it. These two patterns give a simple placement rule: a decoy must either sit on a sweep path or imitate a sought asset. A decoy that does neither is unlikely to be reached, however convincing it is.

The downstream scores also show what kind of signal those decoys produce. Intelligence Yield is rarely the limiting factor: most decoy interactions can reveal useful information. The harder questions are whether the attacker is likely to interact with the decoy, and whether that interaction clearly separates attacker activity from legitimate activity. The extremes show a practical trade-off. Broad coverage comes from decoys in busy places, where hits need more interpretation. High Malice Fidelity comes from decoys in places legitimate users should not touch, but those opportunities are fewer.

We provide the rubric, decision rules, and full per-technique assessment so the map can be reproduced, challenged, and updated as ATT\&CK evolves. The result is not a deployment recipe, but a boundary map: it shows where infrastructure deception can start, where it cannot, and what kind of evidence a decoy interaction is likely to produce.

\section*{Artifact Availability}

The rubric, decision rules, and per-technique assessment are anonymized and available at the following temporary URL: \url{https://brilliant-taffy-73a236.netlify.app/}.

\section*{Ethical Considerations}
This evaluation was conducted with approval of the faculty ethics board. The expert study used verbal consent for the recording of the interviews and anonymized, aggregated reporting. Otter.ai was used to record and transcribe interviews. Claude Code Opus 4.8 was used as note taker during the scoring sessions, not as scorer. The released rubric, decision rules, and rationales are authored entirely by the authors. Claude Opus 4.8 was used for grammar and flow checks during writing this manuscript. The authors take full responsibility for the content of the published article.

\section*{Acknowledgements}
[Acknowledgements and funding information have been removed to maintain the integrity of the double-blind review process. They will be restored in the final version.]


\bibliographystyle{splncs04}
\bibliography{references,web-references}


\appendix

\section{Asset Inventory and Decoy-Targetable Assets}
\label{app:assets}

This appendix lists example client asset types identified during technique scoring, along with example decoys and example attacker interactions that can trigger a signal (Table~\ref{tab:asset-taxonomy}); the inventory is not intended to be exhaustive.

\begin{table*}[!htbp]
\centering
\caption{Example client asset types identified during technique scoring, with example decoys and example triggers (attacker actions that can fire the decoy). Asset classes are grouped by the broad layer they belong to.}
\label{tab:asset-taxonomy}
\begingroup
\scriptsize
\setlength{\tabcolsep}{3.2pt}
\renewcommand{\arraystretch}{1.15}
\begin{tabularx}{\textwidth}{>{\raggedright\arraybackslash}p{3.1cm} >{\raggedright\arraybackslash}p{3.1cm} >{\raggedright\arraybackslash}X}
\toprule
\textbf{Client Asset} & \textbf{Example Decoy} & \textbf{Example Trigger} \\
\midrule
\multicolumn{3}{l}{\textit{Identity and credentials}} \\
User / Service Account     & Honey Account            & enumerate, authenticate, access \\
Stored Credential          & Honey Credential         & read, copy \\
Authentication Material     & Honey Token / Ticket     & use (authentication event) \\
Public Secret              & Honey Secret             & use (API call / auth attempt) \\
Process / Memory Region    & Honey Process            & memory read, credential dump \\
Browser Profile            & Honey Browser Profile    & read profile, extract passwords \\
\midrule
\multicolumn{3}{l}{\textit{Data and storage}} \\
File / Document            & Honey File               & read, write, execute \\
Network Share              & Honey Share              & enumerate, read, write \\
Database                   & Honey Database           & connect, query, dump \\
Code Repository            & Honey Repository         & clone, browse, credential extraction \\
Backup / Snapshot          & Honey Backup             & access, copy, mount \\
Removable Media            & Honey USB                & insertion, file access \\
\midrule
\multicolumn{3}{l}{\textit{Endpoints, services, and network}} \\
Network Service            & Honey Service            & probe, auth attempt, exploit \\
Host                       & Honey Host               & land, enumerate, move \\
Network Device             & Honey Network Device     & probe, SNMP query, config dump \\
DNS Record                 & Honey DNS Record         & resolution \\
Network Traffic            & Honey Credentials in Transit & capture and use \\
Broadcast / Multicast      & Honey Broadcast Responder & resolve, poison attempt \\
Wireless Access Point      & Honey AP                 & connect, auth attempt \\
Printer / IoT Device       & Honey IoT Device         & probe, connect, admin access \\
\midrule
\multicolumn{3}{l}{\textit{Cloud, container, and virtualization}} \\
Cloud Resource             & Honey Cloud Object       & enumerate, read, authenticate \\
Cloud Application          & Honey OAuth App          & enumerate, consent, use secret \\
Container                  & Honey Container          & enumerate, execute, escape \\
Hypervisor                 & Honey Hypervisor         & probe, API query, admin command \\
Serverless Function        & Honey Serverless Function & invocation, env-var access \\
CI/CD Pipeline             & Honey Pipeline           & trigger build, access secrets \\
Software Package           & Honey Package            & install, import, execute \\
\midrule
\multicolumn{3}{l}{\textit{System, web, and human}} \\
System Artifact            & Honey System Artifact    & enumerate, read \\
Web Property               & Honey Web Property       & visit, form submission, cred entry \\
\bottomrule
\end{tabularx}
\endgroup
\end{table*}

\end{document}